\documentclass[12pt]{article}
\usepackage[hypertex]{hyperref}
\usepackage[dvipdfmx]{graphicx} 
\usepackage{graphicx,amsmath,amssymb,amsfonts,cite,bm}

\def\a{\alpha} \def\b{\beta} \def\g{\gamma} \def\G{\Gamma} \def\d{\delta} \def\D{\Delta} \def\e{\epsilon}      \def\l{\lambda}  \def\m{\mu} \def\n{\nu}      \def\s{\sigma}       

\def\d{{\bm d}}  \def\D{{\bm D}} \def\A{{\bm A}} \def\F{{\bm F}} 

\newcommand{\sla}[1]{#1 \!\!\!/}

\def\dg{\dagger} \def\del{\partial} \def\nn{\nonumber}
\newcommand{\lsp}{ \left ( }
\newcommand{\rsp}{ \right ) }
\newcommand{\Lg}{\mathcal{L}}
\newcommand{\we}{\wedge}
\newcommand{\To}{\Rightarrow}

\newcommand{\vev}[1]{ \langle {#1} \rangle }

\newcommand{\tr}{{\rm tr}}
\newcommand{\Tr}{{\rm Tr}}

\newcommand{\row}[2]{ \begin{pmatrix}  #1 & #2   \end{pmatrix}  }
\newcommand{\column}[2]{ \begin{pmatrix}  #1 \\ #2 \\  \end{pmatrix} }

\newcommand{\diag}[2]{ \begin{pmatrix}  #1 & 0 \\ 0 & #2 \\   \end{pmatrix}  }

\begin{document}

\begin{titlepage}

\begin{flushright}
KEK-TH-1870
\end{flushright}

\vskip 1.35cm

\begin{center}
{\LARGE \bf Reconstruction of the standard model \\ with classical conformal invariance \\[10pt] in noncommutative geometry}

\vskip 1.2cm

Masaki J.S. Yang

\vskip 0.4cm

{\it Institute of Particle and Nuclear Studies,\\
High Energy Accelerator Research Organization (KEK)\\
Tsukuba 305-0801, Japan\\
}


\begin{abstract} 

In this paper, we derive the standard model with classical conformal invariance 
from the Yang--Mills--Higgs model in noncommutative geometry (NCG). 
In the ordinary context of the NCG,  the {\it distance matrix} $M_{nm}$ which corresponds to the 
vacuum expectation value of Higgs fields is taken to be finite. 
However, since $M_{nm}$ is arbitrary in this formulation, we can take all $M_{nm}$ to be zero. 
In the original composite scheme, the Higgs field itself vanishes with the condition  $M_{nm} = 0$. 
Then, we adopt the elemental scheme, in which the gauge and the Higgs bosons are regarded as elemental fields. 
By these assumptions, all scalars do not have vevs at tree level. The symmetry breaking mechanism will be implemented by the Coleman--Weinberg mechanism.  

As a result, we show a possibility to solve the hierarchy problem in the context of NCG.
Unfortunately, the Coleman--Weinberg mechanism does not work in the SM Higgs sector, 
because the Coleman--Weinberg effective potential becomes unbounded from below for $m_{t} > m_{Z}$.
However, viable models can be possible by proper extensions.

\end{abstract} 

\end{center}
\end{titlepage}

\section{Introduction}

%
The existence of the Higgs boson \cite{Aad:2012tfa,Chatrchyan:2012ufa} 
ask us further questions: {\it e.g.,} its theoretical origin, and the hierarchy problem \cite{tHooft:1979bh}.
Among various approaches which try to explain the origin of the  boson, 
the Yang--Mills--Higgs model in noncommutative geometry (NCG) \cite{Connes:1990qp} is an elegant possibility. 
It treats the Higgs boson as a gauge boson along the fifth dimension 
which has a noncommutative differential algebra.
This class of models has been explored eagerly in this two decades 
\cite{Chamseddine:1992kv, Chamseddine:1992nx, Connes:1994yd, Chamseddine:1996zu, Lizzi:2000bc, Stephan:2005uj, Chamseddine:2006ep, Chamseddine:2013rta, Devastatao:2014xga,Yang:2015gsa,Yang:2015xta}. 
However, these models still require fine tunings in the Higgs potential, and do not solve the hierarchy problem.

Meanwhile, several solutions of the hierarchy problem are proposed in the phenomenological region, 
such as supersymmetry \cite{Haber:1984rc}, composite Higgs model \cite{Georgi:1984af,Georgi:1984ef,Agashe:2004rs,Contino:2010rs}, extended standard model (SM) with classical conformal invariance \cite{Hempfling:1996ht,Meissner:2006zh,Holthausen:2009uc,Iso:2009ss,Iso:2009nw,Hamada:2012bp,Iso:2012jn,Englert:2013gz}, and so on.
The last class of models based on the Bardeen's argument \cite{Bardeen:1995kv}.
It states that if we impose the classical conformal invariance which is broken only by quantum anomalies on the SM, 
it can be free from the quadratic divergences and solve the hierarchy problem.

In this paper, we derive the standard model with classical conformal invariance 
from the Yang--Mills--Higgs model in NCG. 
In the ordinary context of the NCG,  the {\it distance matrix} $M_{nm}$ which corresponds to the 
vacuum expectation value (vev) of Higgs fields is taken to be finite. 
However, since $M_{nm}$ is arbitrary in this formulation, we can take all $M_{nm}$ to be zero. 
In the original composite scheme, the Higgs field itself vanishes with the condition  $M_{nm} = 0$. 
Then, we adopt the elemental scheme, in which the gauge and the Higgs bosons are regarded as elemental fields. 
It should be emphasized that 
we can not interpret finite $M_{nm}$ as vevs of the Higgs scalars, 
in the elemental scheme with the matrix formalization. 
Since the extended curvature $F_{nm}$ is not written in only $(M + H)_{nm}$, 
we can not identify $(M + H)_{nm} \equiv \Phi_{nm}$ 
as the Higgs fields with vevs, like in the composite scheme.
By these assumptions, all scalars do not have vevs at tree level. The symmetry breaking mechanism will be implemented by the Coleman--Weinberg mechanism \cite{Coleman:1973jx}.  

As a result, we show a possibility to solve the hierarchy problem in this context of NCG.
Unfortunately, the Coleman--Weinberg mechanism does not work in the SM Higgs sector, 
because the Coleman--Weinberg effective potential becomes unbounded from below for $m_{t} > m_{Z}$ \cite{Fujikawa:1978ru}.
However, viable models can be possible by proper extensions such as in Refs. \cite{Hempfling:1996ht,Meissner:2006zh,Holthausen:2009uc,Iso:2009ss,Iso:2009nw,Hamada:2012bp,Iso:2012jn,Englert:2013gz}. We leave it for our future work.

Recent observation shows that the self coupling of the Higgs boson is very close to the critical value, $\l (M_{\rm Pl}) \simeq 0$ \cite{Degrassi:2012ry,Buttazzo:2013uya}. Although the self coupling is finite in this model, the zero self coupling $\l (M_{\rm Pl}) = 0$ is achieved by imposing the condition $dy_{n} \we dy_{m} = 0$. 
It might suggest that the Higgs boson is a remnant of some noncommutative theory at the Planck scale.

This paper is organized as follows. 
In the next section, we review the Higgs model in NCG.
In Sect. 3, we present the reconstruction of the standard model.
Section 4 is devoted to conclusions.

\section{Generalized gauge theory with ordinary exterior derivative in $M^{4} \times Z_{N}$}

Here, we define several definitions and main formalization in the generalized gauge theory with ordinary exterior derivative. 
The following construction of theory is basically based on the formalization by Morita and Okumura \cite{Okumura:1996ez}.

In the $N$-sheeted Minkowski space $M^{4} \times Z_{N}$, 
the coordinates are denoted by $(x^{\m}, n = 1-N)$.
In this setup, the exterior derivative is enlarged as $\d = d + d_{5}$ as follows:
\begin{align}
 \d f \equiv df + d_{5} f \equiv \del_{\m} f dx^{\m} + [M, f] dy, 
\end{align}
where $f$ is an arbitrary matrices $f = f_{nm}$. The differential form $dy$ is dependent to $n$: $dy = {\rm Diag} (dy_{1}, dy_{2}, \cdots , dy_{N})$.
Basically, ``zero (one)-form'' of $dy$ is represented by diagonal (off-diagonal) matrix $f = f_{n} \delta_{nm} ~ (f = f_{nm}, f_{nn} = 0)$. 
$M_{nm}^{\dg} = M_{mn}$ with $M_{nn} = 0$ is the matrix corresponds to the typical scales of the discrete spaces and determines the pattern of the symmetry breakings.  
Since $M_{nm}$ are arbitrary parameters, the formulation still works when $M_{nm} = 0$.
In this case, the Higgs boson has no vevs at tree level, as shown later.
From now on, $M = 0$ and $\d = d$ is assumed. Then the nilpotency of $\d$ is obvious.

As the wedge products of differential $dx^{\m}$ and $dy_{n}$ \cite{Okumura:1996ez},
\begin{align}
dx^{\m} \we dx^{\n} = - dx^{\n} \we dx^{\m}, ~~~
dx^{\m} \we d y_{n} = - d y_{n} \we dx^{\m}, ~~~
dy_{m} \we dy_{n} \neq dy_{n} \we dy_{m} \neq 0, \label{2}
\end{align}
are assumed. 
This $dy_{n}$ does not have the noncommutative property, $dy_{n} f_{nm} = f_{nm} dy_{n}$, 
because the noncommutative differential algebra is undertaken by the matrix algebra, 
such as in many papers \cite{Chamseddine:1992kv, Okumura:1996ez}. 

\vspace{12pt}

In the original paper by Connes and Lott \cite{Connes:1990qp},  and its successors \cite{Chamseddine:1992kv, Chamseddine:1992nx, Chamseddine:1996zu} treat the gauge and Higgs bosons as some kind of composite fields. 
In this picture, a gauge (Higgs) field consists of ``constituent fields'' $a^{i}_{n} (x), b^{i}_{n}(x)$:
\begin{align}
\A_{nm} (x) &= \sum_{i} a^{i}_{n} (x) \d_{nm} b^{i}_{m} (x) \equiv A_{n} (x) \delta_{nm} + \Phi_{nm} (x) dy, \label{4}  \\
A_{n} (x) &= \sum_{i} a_{n}^{i} (x) d b_{n}^{i} (x) , ~~
\Phi_{nm} (x) = \sum_{i} a_{n}^{i} (x) M_{nm} b_{m}^{i} (x) - M_{nm} .
\label{5}
\end{align}
Although this formulation is conceptually interesting, 
the detailed dynamics of the binding mechanism in Eqs.~(\ref{4}) and (\ref{5}) is not specified.
Indeed, this kind of composite theory has been explored in the induced gauge theory \cite{Bjorken:1963vg,Terazawa:1976cx,Terazawa:1976xx}, 
preon models \cite{Shupe:1979fv,Harari:1979gi,Fritzsch:1981zh}, composite Higgs models \cite{Georgi:1984af,Georgi:1984ef,Agashe:2004rs,Contino:2010rs}, and so on.
However, it is highly nontrivial whether the system of the gauge and Higgs bosons (\ref{4}) and (\ref{5}) is phenomenologically viable. 
Additionally, as we can be seen from Eq.~(\ref{5}), the scalar field $\Phi (x)$ vanishes 
$\Phi (x) \to 0$, in the limit of $M_{nm} \to 0$. 
Then, we can not write down a theory of the scalar fields {\it without} vevs in this composite scheme.

For these reasons, it seems to be more natural to regard the gauge and Higgs bosons as elemental fields, 
like several papers \cite{DuboisViolette:1988ir, Coquereaux:1990ev, Sitarz:1993zf, Morita:1993zv}.
The elemental generalized connection $\A (x)$ is introduced as
\begin{align}
 \A(x) &= \A_M(x) dx^M = A_{\m} (x) dx^{\m} + H (x) dy = A(x) + H (x) dy .
\end{align}
Henceforth, we often omit the argument $x$ if there is no confusion. 
In the matrix forms, 
\begin{align}
\A = 
\begin{pmatrix}
A_{1} & H_{12} dy_{2} & \cdots & H_{1N} dy_{N} \\
H_{21} dy_{1} & A_{2} & \cdots & H_{2N} dy_{N} \\
\vdots & \vdots & \ddots & \vdots \\
H_{N1} dy_{1} & H_{N2} dy_{2} & \cdots & A_{N} 
\end{pmatrix} ,
\end{align}
or in components,
\begin{align}
 \A_{nm} (x) &= A_{n \, \m} (x) \delta_{nm} dx^{\m} + H_{nm} (x)  dy_{m} .
\end{align}
Here, $A_{n}$ is the $k_{n} \times k_{n}$ matrix-valued gauge bosons of some gauge groups $\rm G_{n}^{1} \times \cdots \times G_{n}^{r}$. Then, the scalar filed $H_{nm}$ is $k_{n} \times k_{m}$ matrix. 

The covariant exterior derivative is defined by
\begin{align}
\D = \d + \A = (\del_{\m} + A_{n \, \m} )dx^{\m} + H dy , 
\label{covder}
\end{align}
and requiring $\D' G = G \D$ with $G_{nm} = G_{n} (x) \delta_{nm}$, 
the gauge transformation of $\A$ is found to be
\begin{align}
\A' &= G \A G^{-1}- (\d G) G^{-1}  \\
&= ( G A_{\m} dx^{\m} + G H dy - \del_{\m} G dx^{\m}) G^{-1} \\
&= A'_{\m} dx^{\m} + H' dy .
\end{align}
In components,
\begin{equation}
A_{n \, \m}' = G_{n} A_{n \, \m} G_{n}^{-1}  - (\del_{\m} G_{n}) G_{n}^{-1} , ~~~ 
H'_{nm} = G_{nk} H_{kl} G^{-1}_{lm} .
\end{equation}
Then, $A_{n \, \m} (H_{nm})$ transforms as a usual gauge (bi-fundamental scalar) field. 
The gauge transformation of the fifth component becomes an incomplete form because $d_{5} = 0$\footnote{Perhaps one might suspect that the fifth gauge boson does not associate with the fifth derivative operator. However,  it might be interpreted as some kind of remnant zero mode, which usually occur in a theory with normal extra dimension.}. 
Note that the gauge invariance does not forbid the mass term of $H_{nm}$. 

We impose the following Hermite condition to the connection, for the sake of the Hermiticity of the Lagrangian, 
\begin{align}
 \A^{\dagger} = -\A, ~~ \To ~~ A_{n \, \m}^{\dg} = - A_{n \, \m}, ~~~ H_{nm}^{\dg} = - H_{mn} .
\end{align}
The field-strength two-form $F$ is defined as a usual form:
\begin{align}
 \F &= \d\A + \A \we \A = (d A + d H dy) + (A + H dy) \we (A + H dy) \\ 
 & = dA + A\we A + (\del_{\m} H + A_{\m} H - H A_{\m} ) \, dx^{\m} \we dy + Hdy \we  H dy ,
\end{align}
or, in components, 
\begin{align*}
 \F_{nm}  = (dA_{n} + A_{n}\we A_{n} )\delta_{nm} + (\del_{\m} H_{nm} + A_{n \, \m} H_{nm} - H_{nm} A_{m \, \m} ) \, dx^{\m} \we dy_{m } + H_{nl} H_{lm} dy_{l} \we dy_{m} .
\end{align*}

It should be emphasized that we can not interpret finite $M_{nm}$ as vevs of the Higgs scalars, 
in the elemental scheme with the matrix formalization. 
The reason is as follows.
When we calculate the extended curvature $\F$ with finite $M_{nm}$, 
Higgs interaction terms proportional to $dy_{l} \we dy_{m}$ are found to be
\begin{align}
F_{nm} &\ni (M_{nl} H_{lm} - {H_{nl} M_{lm}} +  
 H_{nl} H_{lm} ) \, dy_{l} \we dy_{m} \\
&= [M_{nl} M_{lm} - (M + H)_{nl} (M - H)_{lm}] \, dy_{l} \we dy_{m} \label{novevhiggs}.
\end{align}
Since this curvature is not written in only $(M + H)_{nm}$, 
we can not identify $(M + H)_{nm} \equiv \Phi_{nm}$ 
as the Higgs fields with vevs, like in the composite scheme.

In order to build the gauge-invariant Lagrangian, the inner products of two-forms are calculated to be\footnote{In Eq.~(\ref{28}), the second $\delta_{nk} \delta_{ml}$ term vanishes because $dy_{n} \we dy_{m} \neq dy_{m} \we dy_{n}$ is assumed in Eq.~(\ref{2}).} \cite{Okumura:1996ez}
\begin{align}
\vev{dx^\mu \we dx^\nu, dx^\rho \we dx^\sigma} &=g^{\mu\rho}g^{\nu\sigma}- g^{\mu\sigma}g^{\nu\rho}, \label{26}\\
\vev{dx^\mu \we dy_{n}, dx^\nu \we dy_{m}} &=-\a^2 \delta_{nm} g^{\mu\nu} , \label{27}\\
\vev{dy_{n} \we dy_{m}, dy_{l} \we dy_{k}}&= \a^{4} \delta_{nl} \delta_{mk}  \label{28},
\end{align}
while other products between the basis two-forms to be vanish.

Summarizing these results, the bosonic Lagrangian is described as
\begin{align}
\Lg_{B} &= - {\rm Tr} \vev{\F^{\dg} , S \F} = - \sum_{n,m} \vev{\F_{nm}^{\dg},  S_{n} \F_{nm}} ,
\label{L0}
\end{align}
where $S = {\rm diag} (g_{1}^{-2} E_{1} , g_{2}^{-2} E_{2} , \cdots g_{N}^{-2} E_{N})$, 
$g_{n}$ and $E_{n}$ is the gauge coupling and the unit matrix of the $n$th gauge fields $A_{n}$.
Tr denotes the trace over both external linear space with $n,m = 1-N$ and internal gauge space. 
In components, 
\begin{align}
\Lg_{B} =  \sum_{n} {1\over g_{n}^{2}} \tr \left[ - {1\over 2}  F_{n \, \m\n}^{\dg} F^{\m\n}_{n} 
+  \a^{2} \sum_{m}  | D_{\m} H_{nm} |^{2}  
- \a^{4} \sum_{l} | H_{nl} H_{lm}|^{2} \right ] ,
\end{align}
where $F_{n \, \mu\nu} = \lsp \partial_{\mu} A_{n \,\nu} - \partial_{\nu} A_{n \, \mu} + [A_{n \, \mu}, A_{n \, \nu} ] \rsp, $
and tr denotes the trace over internal gauge spaces.  
By rescaling the boson fields 
\begin{align}
A_{n \, \m} \to i g_{n} A_{n \, \m} ,  ~~~ H_{nm} \to { g_{n} g_{m} \over \a \sqrt{g_{n}^{2} + g_{m}^{2}}  } \, i  H_{nm} ~ (n < m), 
\label{rescale}
\end{align}
we obtain the final Lagrangian with canonical kinetic terms and the following self interaction term:
\begin{align}
V(H) = \sum_{n,m,l}  { g_{l}^{4} (g_{n}^{2} + g_{m}^{2}) \over (g_{n}^{2} + g_{l}^{2}) (g_{l}^{2} + g_{m}^{2})  }  | H_{nl} H_{lm}|^{2} .
\end{align}
Then, in this scheme, the self couplings of the Higgs scalars are comparable to square of the gauge couplings.

The fermionic Lagrangian is also constructed from the generalized connection.  
The Dirac operator in this space is produced by replacing the basis forms $dx^{M}$ in the covariant derivative (\ref{covder})  to the gamma matrices $\G^{M}$:\footnote{The original paper by Okumura construct the Dirac operator by introducing proper inner products like Eqs.~(\ref{26}-\ref{28}). }
\begin{align}
\D &= \d + \A =  (\del_{\m} + A_{n \, \m} )dx^{\m} + H dy , \label{dirac1} \\
\D \!\!\!\!/ \, &= D_{M} \G^{M} =  (\del_{\m} + A_{n \, \m} ) \g^{\m} + H i \g^{5} ,
\end{align}
where $\G^{M} = (\g^{\m}, i\g^{5})$ satisfies the Clifford algebra $\{ \G^{M}, \G^{N} \} = 2g^{MN}$.  

The fermion fields, assigned in each $n = 1-N$, are represented by row and column vector
\begin{align}
\Psi = ( \psi_{1}, \psi_{2}, \cdots, \psi_{N})^{T} , ~~~
\bar \Psi = (\bar \psi_{1}, \bar \psi_{2}, \cdots, \bar \psi_{N}) ,
\end{align}
and the fermionic Lagrangian is defined by a bi-linear form:
\begin{align}
\Lg_{F} = \bar \Psi i \D \!\!\!\!/ \, \Psi 
= \sum_{n,m} \bar \psi_{n} i [(\del_{\m} + A_{n \, \m}) \delta_{nm} \g^{\m} + H_{nm} i \g^{5}] \psi_{m} ,
\label{dirac4}
\end{align}
which satisfies $\Lg_{F}^{\dg} = \Lg_{F}$ with the rescaling (\ref{rescale}).

\subsection{$M^{4} \times Z_{2}$ toy model}

As a typical example, we present $M^{4} \times Z_{2}$ toy model. 
For the following connection $\A$,
\begin{align}
\A = A_{\m n} dx^{\m}  + H_{nm} dy = 
\begin{pmatrix}
A_{1} & H_{12} dy_{2}  \\
H_{21} dy_{1} & A_{2}  \\
\end{pmatrix} ,
\end{align}
the gauge transformation is defined by
\begin{align}
\A' = 
\diag{G_{1}}{G_{2}}
\begin{pmatrix}
A_{1} & H_{12} dy_{2} \\
H_{21} dy_{1} & A_{2} \\
\end{pmatrix}
\diag{G_{1}^{-1}}{G_{2}^{-1}}
+ 
\diag{d G_{1} \cdot G_{1}^{-1}}{ d G_{2} \cdot G_{2}^{-1}} .
\end{align}
The field-strength is calculated as
\begin{align}
\F &= \d \A + \A \we \A = 
\begin{pmatrix}
F_{1} + H_{12} H_{21} \, dy_{2} \we dy_{1} & D_{\m} H_{12} \, dx^{\m} \we dy_{2} \\
D_{\m} H_{21} \, dx^{\m} \we dy_{1} & F_{2} + H_{21} H_{12} \, dy_{1} \we dy_{2}
\end{pmatrix} ,
\end{align}
where $F_{n} = d A_{n} + A_{n}\we A_{n}$ and $D_{\m}H_{nm} = \del_{\m} H_{nm} + A_{n} H_{nm} - H_{nm} A_{m}. $

With the rescaling~(\ref{rescale}), the Lagrangian with canonical kinetic terms is found to be
\begin{align}
\Lg_{B} &= - {\rm Tr} \vev{\F^{\dg} , S \F} \\
 &= - {1\over 2} \tr F_{1 \m\n} F_{1}^{\m\n} - {1\over 2} \tr F_{2 \m\n} F_{2}^{\m\n} +
 \tr [ | D_{\m} H | ^{2} - \l | H^{\dg} H |^{2} ] , 
\label{Lag2}
\end{align}
where we rename $H_{12} \to H $. The self-coupling of the Higgs boson is  
$\l = { g_{1}^{2} g_{2}^{2} / (g^{2}_{1} + g_{2}^{2}) }$.

\section{Reconstruction of the standard model without mass scale}

In this section we proceed to the reconstruction of the SM without mass scale in the noncommutative geometry.
For simplicity, we consider only one generation and omit internal flavor space. 
The introduction of the Yukawa interaction is found in many literatures, such as in Ref.~\cite{Okumura:1997qi}.
The internal gauge space is assumed to be eight dimension. The identity matrix is represented 
\begin{align}
1_{8} = 1_{2} \otimes 1_{4}, 
\end{align}
where $1_{n}$ is the identity matrix of the $n$ dimensional space. 
Each subspaces are correspond to those of ${\rm SU(2)}_{L,R}$ and ${\rm SU(3)}_{c} \times {\rm U(1)}_{B-L}$ respectively. 

In this space, the elemental extended connection is defined in $2 \times 8$ = 16 dimensional space (we rename the index of the discrete space $(1,2) \to (L,R)$):
\begin{align}
\A = A_{\m}(x,y) dx^{\m} +  A_{5}(x,y) dy = 
\begin{pmatrix}
A_{L} & H dy_{R}  \\ H^{\dg} dy_{L} & A_{R}
\end{pmatrix} .
\end{align}
Here,
\begin{align}
A_{L \, \m}(x) &= -{i \over 2} \s^{a} A^{a}_{\m} (x) \otimes 1_{4} - { i \over 2} Y_{L} B_{\m}(x)  - {i \over 2}1_{2} \otimes \l'{}^{\a}  G^{\a}_{\m}(x) , \label{39}  \\
A_{R \, \m} (x)&=  - { i \over 2} Y_{R} B_{\m}(x)  - {i \over 2}  1_{2} \otimes \l'^{\a} G^{\a}_{\m}(x),  \label{40} 
\end{align}
where $\s^{a}$ are the Pauli matrices with $a=1-3$ and $\l'^{\a}$ are the generator of SU(3) with $\a = 1-8$, which is embedded in the $4 \times 4$ representation space as follows:
\begin{align}
\l'{}^{\a} = 
\begin{pmatrix}
 & & & 0 \\
 & \l^{\a} & & 0 \\
 & & & 0 \\
0 & 0 & 0 & 0
\end{pmatrix} .
\end{align}
Here, $\l^{\a}$ is the Gell-Mann matrices which satisfies ${\rm tr} [\l^{\a} \l^{\b}] = 2 \delta^{\a\b}$.
The hypercharges $Y_{L,R}$ are
\begin{align}
& Y_{L} 
= ({1\over3}, {1\over3}, {1\over3}, -{1},{1\over3}, {1\over3}, {1\over3}, -{1}), \\
& Y_{R} 
=  ({4\over3}, {4\over3}, {4\over3}, 0, -{2\over3}, -{2\over3}, -{2\over3}, -2).
\end{align}
%
The gauge transformation matrix $G = {\rm diag} (G_{L}, G_{R})$ is represented by
\begin{align}
G_{L} &= \exp[i \s^{a} \a^{a} \otimes 1_{4}]  \cdot \exp[i Y_{L} \b] \cdot \exp[i 1_{2} \otimes \l'{}^{\a} \g^{\a} ] , \label{g1} \\
G_{R} &= \exp[i Y_{R} \b] \cdot \exp[i 1_{2} \otimes \l'{}^{\a} \g^{\a}] , \label{g2}
\end{align}
where $\g^{\a}, \a^{a}, \b, $ are gauge transformation function of the SM gauge groups ${\rm SU(3)}_{c} \times {\rm SU(2)}_{L}\times {\rm U(1)}_{Y}$ respectively. 
From Eqs.~(\ref{g1}) and (\ref{g2}), the gauge transformation property of $H$ is determined 
\begin{align}
H ' &= G_{L} H G_{R}^{-1} \\
&= \exp[i \s^{a} \a^{a} \otimes 1_{4}] \exp[i 1_{2} \otimes \l'{}^{\a} \g^{\a} ] 
~ H ~ 
\exp[- i 1_{2} \otimes \l'{}^{\a} \g^{\a}] \exp[- i \s^{3} \b \otimes 1_{4}] .
\end{align}
Then, generally $H$ transforms as $({\bf 1+8}, {\bf 2}, \pm 1/2)$ representation under the gauge groups of the SM.
This Higgs model inspired NCG (including the original composite scheme) allow a color-octet Higgs scalar. 
Phenomenologically it is interesting possibility and several author discussing on such scalar bosons \cite{Manohar:2006ga, Cao:2013wqa}, we exclude this alternative. 

Imposing the following constraint, 
\begin{align}
H = h \otimes 1_{4},  
\end{align}
with a $2 \times 2$ matrix $h$, the gauge transformation of the Higgs boson becomes
\begin{align}
H ' &= G_{1} H G_{2}^{-1} = \exp[i \s^{a} \a^{a} \otimes 1_{4}] \, H \, \exp[- i \s^{3} \b \otimes 1_{4}] .
\end{align}
Then we assign
\begin{align}
H &= 
\begin{pmatrix}
H_{u}^{0} & H_{d}^{+} \\ H_{u}^{-} & H_{d}^{0} 
\end{pmatrix}
\otimes 1_{4}  \equiv (H_{u} , H_{d}) \otimes 1_{4}. 
\end{align}
Therefore, it leads to some restricted class of the two Higgs doublet models (2HDM) \cite{Lee:1973iz}. 
If we impose an additional constraint $H_{u} = \tilde H_{d}$, where $\tilde H = i \s^{2} H^{*}$, 
it result in the SM which has only one Higgs doublet. 

\vspace{12pt}

When the connection $A_{n}$ is a direct sum of multiple gauge bosons like (\ref{39}), (\ref{40}),
there are several methods to assign the different gauge couplings. 
Here, we assume the gauge coupling matrix $S$ is dependent to the internal space as follows:
\begin{align}
S &= {\rm diag} (S_{L}, S_{R}) ,  \\
S_{L,R} &= 1_{2} \otimes ({1\over 3} a_{L,R}, {1\over 3} a_{L,R}, {1\over 3} a_{L,R}, b_{L,R}) \label{48} .
\end{align}
From the Lagrangian~(\ref{L0}) with rescaling of the connections,  the bosonic Lagrangian is calculated as
\begin{align}
\Lg_{B} = &- {1\over 4} F_{\m\n}^{a} F^{a \m\n} - {1\over 4} B_{\m\n} B^{\m\n} - {1\over 4} G_{\m\n}^{\a} G^{\a \m\n} \nn \\
&+ \tr \, (D_{\m} H)^{\dg} (D_{\m}H) - \l \, \tr | H^{\dg} H |^{2} ,
\end{align}
where 
\begin{align}
F_{\m\n}^{a} &= \del_{\m}A_{\n}^{a} - \del_{\n} A_{\m}^{a} + g \e^{abc} A^{b}_{\m} A^{c}_{\n}, \\
B_{\m\n} &= \del_{\m} B_{\n} - \del_{\n} B_{\m}, \\
G_{\m\n}^{a} &= \del_{\m} G_{\n}^{\a} - \del_{\n} G_{\m}^{\a} + g_{c} f^{\a\b\g} G^{\b}_{\m} G^{\g}_{\n}, \\
D_{\m} H &= \del_{\m} - {i \over 2} g \s^{a} A^{a}_{\m} H - {i\over 2} H  g' \s^{3} B_{\m} .
\end{align}
Here, the gauge couplings are found to be
\begin{align}
{1 \over g^{2}} = a_{L} + b_{L}, ~~~ 
{1 \over g'{}^{2}} =  {1\over 9} a_{L} +  b_{L} + {10 \over 9} a_{R}  + 2 b_{R}, ~~~ 
{1 \over g_{c}^{2}} = 2 (a_{L} + a_{R}) ,
\end{align}
and the Higgs self coupling is 
\begin{align}
{1 \over \l} = {a_{L} + b_{L} + a_{R} + b_{R}} .  \label{higgssc}
\end{align}
These couplings are not independent and the following formula holds,
\begin{align}
{4 \over 9} {1\over g_{c}^{2}} + {1\over g^{2}} + {1\over g'{}^{2}} = {2 \over \l} \, .
\end{align}
In principle, we can break this relation taking the gauge coupling matrix $S_{R} = (x_{R}, y_{R}) \otimes ({1\over3} a_{R} 1_{3}, b_{R})$. Indeed this is the most general form of $S_{R}$ which commutes with $A_{R}$.
In any case, the size of the Higgs self coupling is roughly equal to square of the gauge couplings, $\l \sim g^{2}$.

The Higgs potential is found to be 
\begin{align}
V(H_{u}, H_{d}) & = \l |H^{\dg} H| = \l \Tr
\begin{pmatrix}
H_{u}^{\dg} H_{u} & H_{u}^{\dg} H_{d} \\ 
H_{d}^{\dg} H_{u} & H_{d}^{\dg} H_{d}
\end{pmatrix} 
\begin{pmatrix}
H_{u}^{\dg} H_{u} & H_{u}^{\dg} H_{d} \\ 
H_{d}^{\dg} H_{u} & H_{d}^{\dg} H_{d}
\end{pmatrix} 
\nn \\
&= \l ( |H_{u}^{\dg} H_{u}|^{2} + 2 |H_{u}^{\dg} H_{d}|^{2} +  |H_{d}^{\dg} H_{d}|^{2}) .
\end{align}
This is rather restricted form of the general potential of the 2HDM.

\subsection{The fermionic sector}

The construction of the Dirac Lagrangian is achieved by the discussion along Eqs.~(\ref{dirac1})-(\ref{dirac4}). 
In this space, the SM fermions are assigned as
\begin{align}
\psi(x,L) = 
\begin{pmatrix}
{u_{L}} \\ {\n_{L}} \\ {d_{L}} \\ {e_{L}}
\end{pmatrix}
, ~~ 
\psi(x,R) = 
\begin{pmatrix}
{u_{R}} \\ {0} \\ {d_{R}} \\ {e_{R}}  
\end{pmatrix}
. ~~
\end{align}
where the color index is omitted.
The covariant derivative is written as
\begin{align}
\D_{M} =  \d_{M} + \A_{M} = 
\lsp
\diag{\del_{\m} + A_{L\m}}{\del_{\m} + A_{R\m}} ,  
\begin{pmatrix}
0 & H \\ - H^{\dg} & 0 
\end{pmatrix}
\rsp .
\end{align}
By rescaling the connections $A_{L,R} \to i g A_{L,R}$, $H \to i \sqrt{\l} H$, 
 the Dirac Lagrangian (\ref{dirac4}) is found to be
\begin{align}
\Lg_{D} = \bar \Psi i \G^{M} \D_{M} \Psi 
&=  
\row{\bar \psi_{L}}{\bar \psi_{R}} 
\begin{pmatrix}
i \sla{D \!}_{\, L}  & \sqrt{\l} H \\
\sqrt{\l} H^{\dg} & i \sla{ D \!}_{\, R}
\end{pmatrix} 
\column{\psi_{L}}{\psi_{R}} ,
\end{align}
where $\sla{D \!}_{\, L,R}$ are the covariant derivatives
\begin{align}
\sla{D \!}_{\, L} &= \g^{\m} \del_{\m} \otimes 1_{8} - {i\over 2} ( g \s^{a} A^{a}_{\m} \otimes 
1_{4} + g'  Y_{L} B_{\m} + g_{c} 1_{2} \otimes \l'{}^{\a} G^{\a}_{\m} ) , \\
\sla{D \!}_{\, R} &= \g^{\m} \del_{\m} \otimes 1_{8} - {i\over 2} ( g' Y_{R} B_{\m} + g_{c} 1_{2} \otimes \l'{}^{\a} G^{\a}_{\m} ) .
\end{align}
Then, we found that the Yukawa coupling is related to the self coupling of the Higgs $y = \sqrt{\l} \sim g$ in this toy model.

\section{Conclusions}

In this paper, we derived the standard model with classical conformal invariance 
from the Yang--Mills--Higgs model in NCG. 
In the ordinary context of the NCG,  the {\it distance matrix} $M_{nm}$ which corresponds to the 
vacuum expectation value (vev) of Higgs fields is taken to be finite. 
However, since $M_{nm}$ is arbitrary in this formulation, we can take all $M_{nm}$ to be zero. 
In the original composite scheme, the Higgs field itself vanishes with the condition  $M_{nm} = 0$. 
Then, we adopt the elemental scheme, in which the gauge and the Higgs bosons are regarded as elemental fields. 
It should be emphasized that 
we can not interpret finite $M_{nm}$ as vevs of the Higgs scalars, 
in the elemental scheme with the matrix formalization. 
Since the extended curvature $F_{nm}$ (\ref{novevhiggs}) is not written in only $(M + H)_{nm}$, 
we can not identify $(M + H)_{nm} \equiv \Phi_{nm}$ 
as the Higgs fields with vevs, like in the composite scheme.
By these assumptions, all scalars do not have vevs at tree level. The symmetry breaking mechanism will be implemented by the Coleman--Weinberg mechanism \cite{Coleman:1973jx}.  

As a result, we  show a possibility to solve the hierarchy problem in this context of NCG.
Unfortunately, the Coleman--Weinberg mechanism does not work in the SM Higgs sector, 
because the Coleman--Weinberg effective potential becomes unbounded from below for $m_{t} > m_{Z}$ \cite{Fujikawa:1978ru}.
However, viable models can be possible by proper extensions such as in Refs. \cite{Hempfling:1996ht,Meissner:2006zh,Holthausen:2009uc,Iso:2009ss,Iso:2009nw,Hamada:2012bp,Iso:2012jn,Englert:2013gz}. We leave it for our future work.

Recent observation shows that the self coupling of the Higgs boson is very close to the critical value, $\l (M_{\rm Pl}) \simeq 0$ \cite{Degrassi:2012ry,Buttazzo:2013uya}. Although the self coupling~(\ref{higgssc}) is finite in this model, the zero self coupling $\l (M_{\rm Pl}) = 0$ is achieved by imposing the condition $dy_{n} \we dy_{m} = 0$. 
It might suggest that the Higgs boson is a remnant of some noncommutative theory at the Planck scale.

\section*{Acknowledgements}

The author would like to appreciate K.-y.~Oda for useful discussions and valuable comments.
This study is supported by the Iwanami Fujukai Foundation, 
and partly supported by the Knight Errantry of the Japan Particle Theory Forum.


\end{document}